# Isolated attosecond free-electron laser based on a sub-cycle driver from hollow capillary fibers

Yaozong Xiao[1, 2, †], Tiandao Chen[1, 3, †], Bo Liu[2], Zhiyuan Huang[3, *], Meng Pang[3], Yuxin Leng[3], Chao Feng[2, *]

[1]*Zhangjiang Laboratory, Shanghai 201210, China*

[2]*Shanghai Advanced Research Institute, Chinese Academy of Sciences, Shanghai 201210, China*

[3]*State Key Laboratory of High Field Laser Physics and CAS Center for Excellence in Ultra-intense Laser Science, Shanghai Institute of Optics and Fine Mechanics (SIOM), Chinese Academy of Sciences (CAS), Shanghai 201800, China*

*Corresponding author. Email: huangzhiyuan@siom.ac.cn (Z. H), fengc@sari.ac.cn (C. F)

[†]These authors contribute equally to the work

**Abstract**

The attosecond light source provides an advanced tool for investigating electron motion using time-resolved-spectroscopy techniques. Isolated attosecond pulses, especially, will significantly advance the study of electron dynamics. However, achieving high-intensity isolated attosecond pulses is still challenging at the present stage. In this paper, we propose a novel scheme for generating high-intensity, isolated attosecond soft X-ray free-electron lasers (FELs) using a mid-infrared (MIR) sub-cycle modulation laser from gas-filled hollow capillary fibers (HCFs). The multi-cycle MIR pulses are first compressed to sub-cycle using a helium-filled HCF with decreasing pressure gradient due to soliton self-compression effect. By utilizing such sub-cycle MIR laser pulse to modulate the electron beam, we can obtain a quasi-isolated current peak, which can then produce an isolated FEL pulse with high signal-to-noise ratio (SNR), naturally synchronizing with the sub-cycle MIR laser pulse. Numerical simulations have been carried out, including the sub-cycle pulse generation, electron beam modulation and FEL radiation processes. The simulation results indicate that an isolated attosecond pulse with wavelength of 1 nm, peak power of ~28 GW, pulse duration of ~600 attoseconds and SNR of ~96.4% can be generated by our proposed method. The numerical results demonstrated here pave a new way for generating the high-intensity isolated attosecond soft X-ray pulse, which may have many applications in nonlinear spectroscopy and atomic-site electronic process.

**Keywords:** Isolated X-ray attosecond pulses, high-intensity free electron lasers, sub-cycle pulses, hollow capillary fibers, soliton self-compression.

## 1. Introduction

The study of ultrafast dynamics of microscopic particles can help people elucidate and understand numerous macroscopic phenomena. Attosecond experimental techniques are crucial for explaining electronic dynamics in quantum systems, which occurs on the subfemtosecond scale. Various attosecond ultrafast measurement techniques developed based on attosecond light sources, for example, attosecond transient absorption [1], attosecond wave-mixing spectroscopy [2] and attosecond transient reflectivity spectroscopy [3], make it possible to deeply study the dynamics of electrons. Since high-harmonic generation (HHG) first experimentally demonstrated the attosecond pulses in 2001 [4], the pulse duration has broken through to 43 attoseconds [5] with the unremitting efforts of researchers. However, the low energy conversion efficiency of HHG in soft X-rays and even shorter wavelengths limits its application.

Developed in parallel with HHG, the free electron laser (FEL) [6-8] offers unique advantages in producing high peak power, fully coherent and short-wavelength ultrashort pulses. Over the past two decades, numerous ultrafast schemes based on Self-Amplified Spontaneous Emission (SASE) FEL [9-18] or externally seeded FEL [19-21] have been proposed and developed. Besides, the ultracompact attosecond FEL scheme [22], based on the plasma-based accelerator and optical undulator, has the potential to become an important research direction in the future. The ultrashort pulses from externally seeded FEL typically exhibit higher stability and shorter pulse duration (tens of attoseconds). However, these methods require complex configurations, and their achievable high photon energy is constrained by the harmonic up-conversion efficiency. Due to its simple configuration and broad photon energy coverage, SASE is currently the primary operating mode of FEL facilities worldwide [23-28]. The enhanced-SASE (ESASE) [10], a representative ultrafast scheme, inherits these advantages. However, limited by the few-cycle laser, the output of the ESASE usually has two satellite peaks, significantly reducing the signal-to-noise ratio (SNR). In order to increase the SNR, methods using two-color modulation lasers in ESASE have been proposed to suppress these satellite peaks but cannot completely eliminate them [14,29]. Subsequently, self-modulation methods were introduced to generate isolated ultrafast pulses [11,12]. Nevertheless, self-modulation cannot synchronize well with external pump lasers, resulting in relative jitter on the order of tens of femtoseconds, which hinders its application in pump-probe experiments.

Hollow capillary fibers (HCFs), with broad transmission band and tightly-confined light fields in its hollow-channel core, have been widely used for ultrashort laser pulse compression and efficient frequency up-conversion [30-39]. One remarkable advantage of gas-filled HCF system is the tunable nonlinearity and dispersion landscape through adjusting gas type and pressure [33]. Due to the soliton-effect self-compression [30,34] in gas-filled HCFs, the laser pulses can be significantly compressed with the combined effect of waveguide-induced anomalous dispersion and self-phase modulation (SPM). Compared with the traditional post compression schemes which can only produce few-cycle pulses [40,41], soliton self-compression holds the capability of generating sub-cycle pulses. As early as 2019, John Travers et al. have observed the sub-cycle self-compression of near-infrared (NIR) pulses in gas-filled HCFs and inferred the waveforms of sub-femtosecond field [30]. In addition, the use of gas-filled HCF with decreasing pressure gradient can further enhance this extreme pulse compression and exhibit high-quality sub-cycle compression [35,36]. The sub-cycle pulses generated by this method can be used for modulating the electron beam and serve as the pump laser that naturally synchronize to the attosecond FEL pulse, thereby solving the long-standing challenges that hinder the scientific application of ESASE.

In this work, we propose a new scheme which combines the advantages of HCF self-compression technique with ESASE for generating intense isolated attosecond pulses. Starting from the conventional mid-infrared (MIR) lasers, in gas-filled HCFs with reducing pressure gradient, the sub-cycle pulses will be produced by extreme soliton self-compression effect and be subsequently used to modulate the electron beams in a wiggler. After energy modulation introduced by the sub-cycle pulses at the output end of HCFs is converted into density modulation, we can obtain an isolated current spike, which plays a crucial role in producing isolated ultrafast FEL radiation. The proposed scheme provides an effective and practical method for generating high-intensity, high-SNR isolated attosecond soft X-ray FELs, which can be used for research in electron dynamics and other related fields.

## 2. Numerical methods

The proposed scheme for high-intensity isolated attosecond FEL pulses generation system includes two stages: sub-cycle MIR pulses generation in the gas-filled HCF and high-intensity, high-SNR attosecond soft X-ray FEL radiation, as illustrated in Fig. 1.

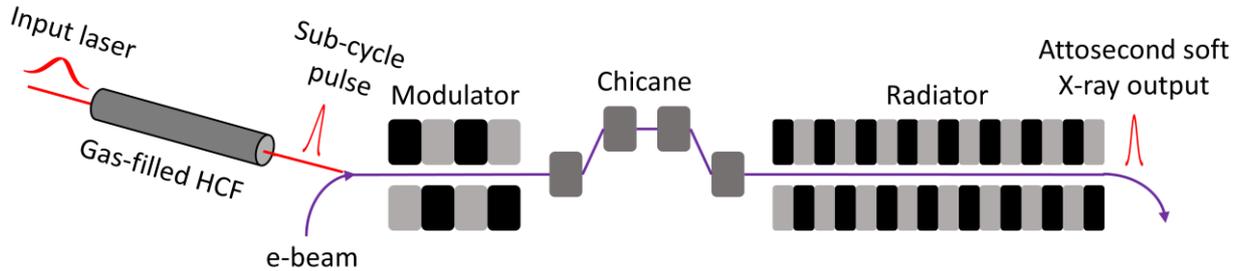

Fig. 1. Layout of attosecond soft X-ray generation driven through MIR sub-cycle pulses from gas-filled HCF.

In the first stage, the MIR laser pulses with multiple optical cycles are launched into the HCF with decreasing pressure gradient and then compressed into sub-cycle range due to soliton self-compression [30,35]. The spatial profile of electrical field inside the HCF is considered to have cylindrical symmetry and can be expressed as a linear superposition of HCF modes [42]:

$\tilde{E}(z,r,\omega) = \sum_j \tilde{E}_j(z,\omega) J_0(u_{1j}\frac{r}{a})$, where $\tilde{E}$ is the electric field in frequency domain, $\omega$ is the angular frequency, $z$ is the propagation distance along HCF, $\tilde{E}_j$ is the electric field at $HE_{1j}$ mode, $J_0$ is the zero order Bessel function, $u_{1j}$ is the $j^{th}$ zero of $J_0$, $r$ is the radial coordinate and $a$ is the core radius of the HCF. The propagation of ultrafast laser pulses composed of multiple modes in gas-filled HCFs can be simulated using the 2D unidirectional pulse propagation equation (UPPE), which can be written as [43]:

$$\frac{\partial \tilde{E}_j(z,\omega)}{\partial z} = i\left(\beta_j(z,\omega) - \frac{\omega}{v_1}\right)\tilde{E}_j(z,\omega) - \frac{\alpha_j(z,\omega)}{2}\tilde{E}_j(z,\omega) + i\frac{\omega^2 \tilde{P}_j^{NL}(z,\omega)}{2c^2\varepsilon_0\beta_j(z,\omega)} \quad (1)$$

$$\tilde{P}_j^{NL}(z,\omega) = 2\pi \int_0^a r\,dr\, J_0(u_{1j}\frac{r}{a})\tilde{P}^{NL}(z,r,\omega) \quad (2)$$

where $v_1$ is the reference frame velocity of the pump pulse at fundamental mode, $c$ is the speed of light in vacuum, $\varepsilon_0$ is the vacuum permittivity, $\alpha_j$ and $\beta_j$ are the fiber loss and propagation constant at $HE_{1j}$ mode, which can be calculated using the Marcatili model [37], and $\tilde{P}^{NL}$ is the nonlinear polarization in frequency domain. Since the gas filled into HCF is a noble gas, the Raman effect can be negligible. The nonlinear polarization includes the Kerr nonlinearity and ionization-induced plasma response, which can be given as [37,38,43]:

$$\tilde{P}^{NL}(z,r,\omega) = F\left[\varepsilon_0\chi^{(3)}E(z,r,t)^3 + P_{ion}(z,r,t)\right] \quad (3)$$

where $F$ represents the time domain Fourier transform, $\chi^{(3)}$ is the third-order nonlinear susceptibility that is relative to Kerr effect, $E$ is the electric field in time domain, $t$ is the time in a reference frame travelling with a group velocity. $P_{ion}$ is the ionization-induced plasma response given by [37,38,43]:

$$\frac{\partial P_{ion}(z,r,t)}{\partial t} = \frac{I_P}{E(z,r,t)}\frac{\partial \rho(z,r,t)}{\partial t} + \frac{e^2}{m_e}\int_{-\infty}^{t}\rho(z,r,t')E(z,r,t')dt' \quad (4)$$

where $I_P$ is ionization potential of the gas, $\rho$ is the plasma density, $e$ is the electronic charge, and $m_e$ is the mass of the electron. The plasma density can be expressed as:

$$\frac{\partial \rho}{\partial t} = W(I)(\rho_{nt} - \rho) + \frac{s}{I_P}\rho I \tag{5}$$

where $W$ is the ionization rate that depends on the intensity of laser pulses, $I$ is the laser pulse intensity, $\rho_{nt}$ is the neutral gas density, and $s$ is the cross section describing the process of collisional ionization.

In the simulation, the electric field can be expressed as $E(z,r,t) = \sum_j J_0(u_{1j}\frac{r}{a})A_j(z,t)\cos(\omega_0 t + \phi_{CEP})$, where $A_j(z,t)$ is the time-dependent complex amplitude (envelope) at $HE_{1j}$ mode, $\omega_0$ is the carrier frequency of the input pulses, and $\phi_{CEP}$ is the carrier-envelope phase (CEP) of the laser pulses. Moreover, the HCF is filled with a gradually decreasing pressure gradient, which can ensure that the incident pulse has sufficient gas broadening spectrum in the HCF, and the sub-cycle pulse at the HCF output port would not be broadened in the time domain due to dispersion caused by gas or other optical elements. Therefore, when the gas pressure at the HCF input port is $p_0$, and the gas pressure at the output port is 0, the pressure distribution along the HCF can be expressed as [44]:

$$p(z) = p_0\sqrt{1 - \frac{z}{L}} \tag{6}$$

where $L$ is the HCF length. We solved the Eq. (1) using a symmetric split-step Fourier method, and calculated the nonlinear term through the fourth-order Runge-Kutta algorithm.

In the second stage, for the purpose of producing the high-intensity attosecond pulses, we adopted an electron beam with high-quality parameters from SwissFEL [45] in our simulation. As sketched in Fig. 1, the sub-cycle MIR pulses generated in the HCF are transmitted into the modulator where they will interact with the electron beams, inducing energy modulation. Afterwards, the electron beam passes through a chicane (dispersion element) downstream of the modulator, and the energy modulation will be transformed into the density modulation, resulting in the generation of an isolated current spike. Finally, the isolated attosecond soft X-ray FELs can be generated in the radiator.

## 3. Results and analysis

### 3.1. Sub-cycle pulses generation in gas-filled HCFs

In the simulation, we used 40 fs (full width half maximum, FWHM), 4 μm, 420 μJ Gaussian-shape MIR pulses as input, as shown in Fig. 2(a). Such high-energy ultrafast MIR pulses are typically obtained through the OPCPA (optical parametric chirped pulse amplification) method. The input MIR pulses are launched in the He-filled HCF using focusing optical elements such as lens, concave and parabolic mirrors, and only the fundamental mode is excited at the HCF input port. The HCF is 30-cm-long with 300-μm core diameter. The HCF input port is filled with a pressure of 113 bar, while the output port is kept as vacuum as possible. This creates a gradually decreasing pressure gradient, which is more conducive to the generation of sub-cycle laser pulses. Figure 2(b) shows the temporal and spectral evolutions of a MIR pulse propagating in a He-filled HCF simulated using the UPPE model. The pulse envelope (brown line), the electric field and pulse spectrum (blue lines) in fundamental mode at three different fiber positions (0, 15 cm, and 30 cm) were also performed in our simulations and the results are plotted in Figs. 2(c)-2(e). It should be noted that in the simulations the energy of the higher-order modes is about two orders of magnitude lower than that of the fundamental mode. Moreover, we found that the peak intensity of pump pulse in the HCF is always far below the photoionization threshold of the He gas, and the ionization fraction of the He gas is always below the level of $10^{-8}$. These results indicate that the spatial nonlinear effect and ionization effect can be neglected in the sub-cycle pulse generation process.

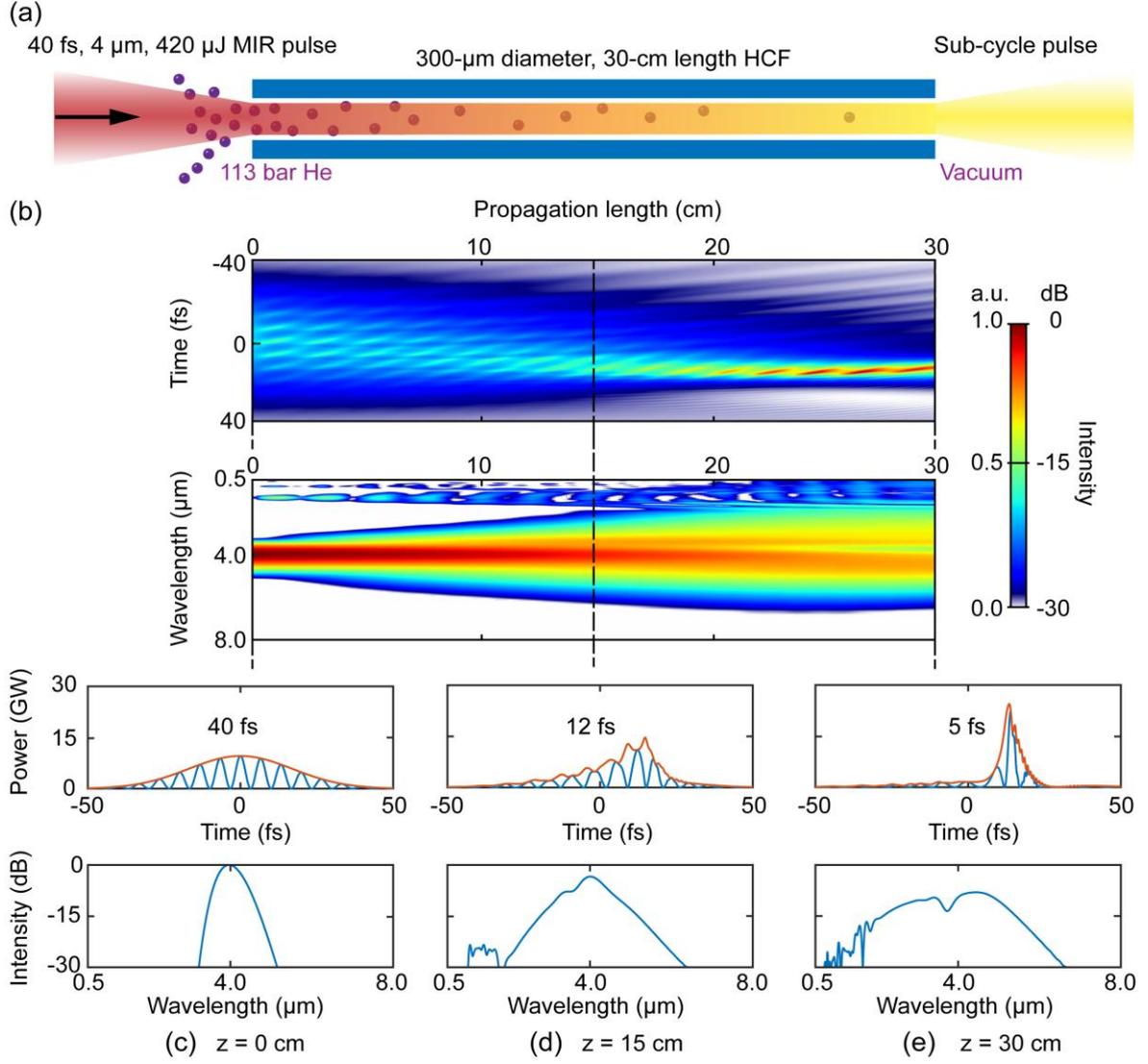

Fig. 2. (a) Conceptual diagram of self-compression process of MIR pulse in He-filled HCFs. (b) Temporal and spectral evolutions of the Gaussian-shape MIR pulse in a 300-μm-diameter He-filled HCF with decreasing pressure gradient (from 113 bar He to vacuum). (c)–(e) The pulse envelope (brown line), the electric field and pulse spectrum (blue lines) in fundamental mode at different propagation positions of (c) z = 0, (d) z = 15 cm, and (e) z = 30 cm.

As shown in Fig. 2(b), the input MIR pulses undergo significant spectral broadening in the first half of the gas-filled HCF. Because of the combined effect of the SPM and anomalous dispersion caused by hollow waveguide, the MIR pulses experience soliton self-compression from an initial 40 fs to 12 fs, as shown in Figs. 2(c) and 2(d). In the latter half of the HCF, the self-compression of the pulses will be further enhanced, as the use of negative pressure gradient can increase the waveguide-induced anomalous dispersion. Therefore, the MIR laser pulses are further compressed from 12 fs to 5 fs, corresponding to optical cycle ranging from ~0.9 to ~0.4, see Figs. 2(d) and 2(e). In the Fig. 2(d), we observed faint spectral lines at short-wavelength edge of the pulse spectrum, which are recognized as the third-harmonic (TH) signal of the pump laser. When the TH signal overlaps with the broadened pulse spectrum, some spectral interference peaks can be observed, as shown in Fig. 2(e). It should be noted that in the simulations the intensity of the TH signal is about three orders of magnitude lower than that of pump pulse. Therefore, the TH generation has little effect on soliton self-compression.

**3.2. 3D simulations for high-intensity isolated FEL pulses generation**

The sub-cycle MIR pulse generated from the gas-filled HCF has an ultrashort pulse width of 5 fs and a peak power of ~30 GW, see Fig. 2(e). We also plotted the electric field and envelope amplitude of the sub-cycle pulses, as well as the spectrum in the linear coordinate, as exhibited at Figs. 3(a) and 3(b). The sub-cycle MIR pulses are used for generating high-intensity isolated attosecond FELs in the second stage. Typical parameters of a soft x-ray FEL facility, as given in table 1, have been adopted in 3D simulations for generating high-intensity isolated FEL pulses.

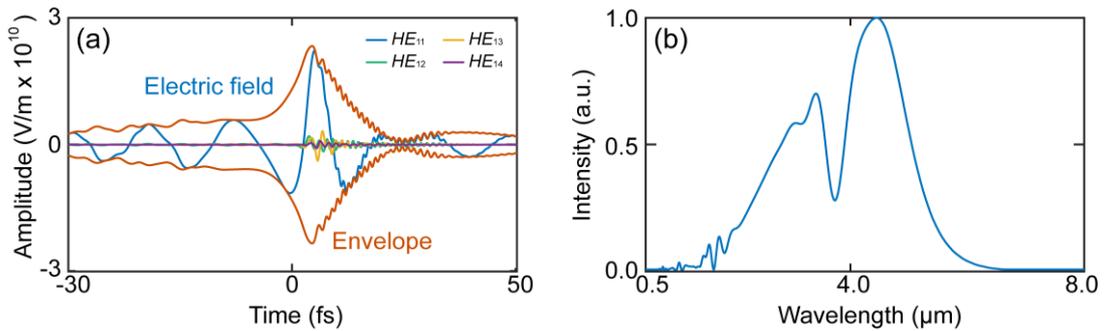

Fig. 3. (a) The electric field (blue) and envelope (brown) amplitude of the sub-cycle pulses in fundamental mode generated at the output port of HCF shown in Fig. 2(e). The electric field in higher-order modes are plot in different colors. (b) The corresponding pulse spectrum.

Table 1. Main parameters in the FEL simulations

| Parameters | Value | Unit |
|---|---|---|
| Electron beam energy | 2500 | MeV |
| Energy spread (RMS) | 0.25 | MeV |
| Normalized emittance | 0.43 | $\mu$mrad |
| Average current | 3000 | A |
| Modulator length (single period) | 0.2 | m |
| Radiator period | 3 | cm |
| Period numbers of each radiator | 100 | / |
| Radiation wavelength | 1 | nm |

We plan to transmit the laser beam to the modulator through free space and concave mirrors. By utilizing appropriate concave mirror and reasonably constructing the beam line, the light field at the focus can be exactly the same as that at the HCF output port [46]. After transmitted into the modulator (see Fig. 1), the sub-cycle MIR pulses generated at the HCF output port, with the transverse size of approximately 0.15 mm root-mean-square (RMS) radius, will interact with the electron beams, and this interaction can lead to a large energy chirp in the electron beams. Then, the energy chirp will be compressed after the electron beam passes through the chicane, resulting in the generation of isolated current spike, as illustrated in Fig. 4. The electron beam has a Gaussian distribution in the transverse direction, and the longitudinal phase space shown here contains the distribution of all electrons in the electron beam. The energy modulation process is implemented using the method described in Ref. [10]. The bunch compression process is simulated with the 3D electron beam tracking code ELEGANT [47], which offers capabilities for modeling collective effects such as intra-beam scattering (IBS), coherent synchrotron radiation (CSR), and incoherent synchrotron radiation (ISR) in accelerators and beam transport systems.

Compared with the fundamental mode in Fig. 3, the contribution of higher-order modes is almost negligible during the electron beam modulation process. Firstly, the electric field intensity of higher-order modes is much weaker than that of the fundamental mode. Secondly, higher-order modes have higher frequencies, which do not satisfy the resonance condition of the undulator: $\lambda_L = \lambda_u (1+K^2/2)/2\gamma^2$, where $\lambda_L$ is the laser wavelength (here $\lambda_L = 4\mu m$), $\lambda_u$ is the modulator period, $K$ is the undulator parameter, and $\gamma$ is electron beam energy normalized by $m_e c^2$ (~0.511 MeV). The resonance condition will select only the fundamental mode for modulation while filtering out other modes. The above two points are also the main reasons why the regions on both sides of the sub-cycle laser cannot produce significant modulation on the electron beam.

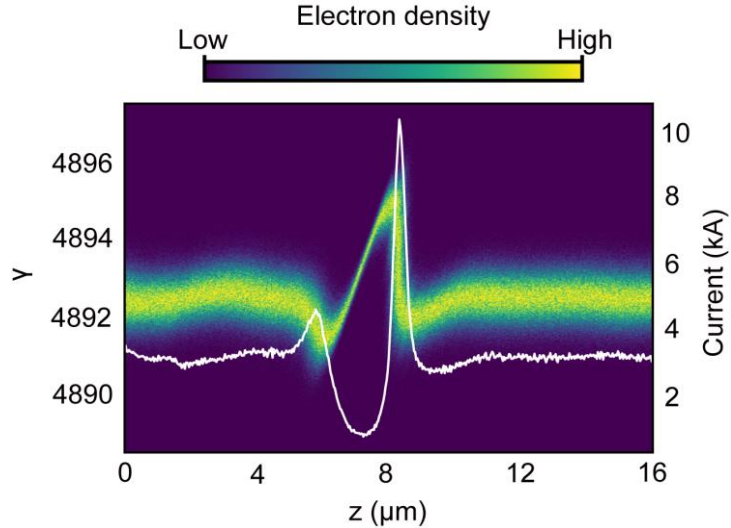

Fig. 4. The energy phase space (color heat map) and the current distribution (white curve) before the radiator.

The current and slice emittance of the electron beam can be well maintained during the transmission process, as demonstrated in multiple ESASE experiments [11,48,49]. However, the longitudinal wakefield caused by the space charge effect cannot be negligible for an electron beam with so high current spike shown in Fig. 4. According to Ref. [14], the longitudinal space charge (LSC) field effectively can be approximated as the result of free space, and we can estimate the LSC field using the following simplified expression:

$$E_s \approx -\frac{Z_0 I'(s)}{4\pi\bar{\gamma}_s^2}\left(2\ln\frac{\bar{\gamma}_s \sigma_s}{r_b} + 1 - \frac{r^2}{r_b^2}\right), \tag{7}$$

where $Z_0 = 377\Omega$ represents free space impedance, $s$ is the longitudinal bunch coordinate, $\bar{\gamma}_s = \gamma/\sqrt{1+K^2/2}$, $I'(s) = dI/ds$ represents the derivative of current with respect to $s$, $\sigma_s$ is the RMS length of the current spike, $r = \sqrt{x^2+y^2}$ and $r_b$ represents the radius of electron beam with uniform transverse distribution. In this work, we take $\gamma = 4892.5$, $K = 1.09$, and $r_b \approx 2\sigma_x \sim 60\mu m$, $\sigma_x$ represents the electron beam RMS transverse size.

The bunch length $\sigma_s$ of peak current shown in Fig. 4 is about 150 nm. As shown in Fig. 5, the electron beam with above parameters would generate the accumulated energy modulation $\Delta E$ (blue lines) after passing through the 17-m undulator. The LSC effect leads to the increase of energy spread, especially the introduction of additional energy chirp at the current peak, corresponding to a peak-to-peak variation of ~15.65 MeV, which will degrade the FEL radiation performance. The FEL amplification process was simulated using GENESIS, a time-dependent 3D FEL code [50]. This powerful tool supports the simulation of single-pass FELs, including SASE FELs, FEL amplifiers, FEL oscillators, and multistage cascaded FELs. The FEL amplification process is modeled by importing the electron beam data, shown in Fig. 4, into GENESIS via a particle file. Additionally, the LSC effect is incorporated into the simulation by importing the beam wakefield file, which is calculated using Eq. (7).

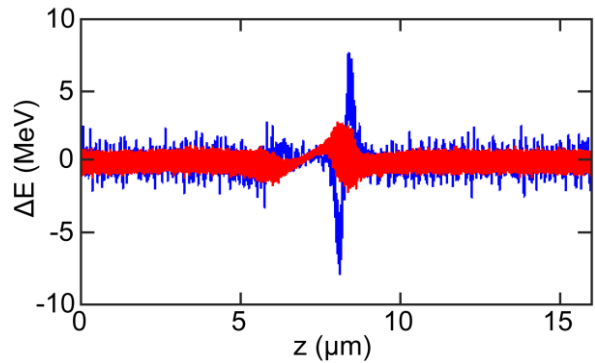

Fig. 5. The energy spread distribution before (red line) and after (blue line) 17-m undulator.

Soft X-rays correspond to the core-level absorption edges of many light elements: carbon (280 eV), nitrogen (410 eV) and oxygen (540 eV), and they are the main elements that make up living organisms. Currently, many attosecond FEL scientific applications are also focused on the soft X-ray range [51,52]. SwissFEL is an internationally renowned soft X-ray free-electron laser facility, with soft X-rays being its advantageous photon energy range. Therefore, in the simulation, we have selected a radiation wavelength in the soft X-ray range, specifically 1 nm, to study the performance of our proposed scheme. For other wavelengths surrounding 1 nm, there will not be a significant difference in the FEL performance.

At the output port of radiator (17 m), the electron beam shown in Fig. 4 can generate an isolated attosecond FEL pulse. In our simulations, the radiation wavelength is chosen as 1 nm. As shown in Fig. 6(a), the ultrafast FEL output pulse has a peak power of ~28 GW, pulse duration of ~600 attoseconds, and an exceptionally high SNR (about 96.4%) compared to a standard ESASE. Additionally, the feature of the sub-cycle pulse generated from gas-filled HCFs would affect the final attosecond soft X-ray FEL radiation. The sub-cycle pulses with different wavelength or waveform will produce different energy modulation, which in turn results in different FEL output. In the Supplementary Materials, we have performed some simulations to study the dependency of attosecond FEL performance and sub-cycle pulse property, and we have exhibited the optimum case among these simulations in this main text. Besides, we numerically investigated how the sub-cycle pulse and the final FEL output would change in the presence of pump energy fluctuations. We assume the pump energy fluctuates by ±1%, and we found the waveform of the generated sub-cycle pulse is nearly unchanged. Accordingly, FEL pulses driven by sub-cycle pulses show tiny change, with only about a 3% fluctuation in peak power. Please see Section VI of the Supplementary Material for more detailed results.

As a comparison, we also simulated the case of the conventional ESASE. We used a 1.5-period few-cycle laser with the central wavelength of 4 μm to modulate the electron beam, resulting in a current modulation with the peak current of ~10.6 kA, which is close to the value in Fig. 4. After passing through a 17 m undulator, such an electron beam can generate an ultrafast FEL pulse with two satellite pulses as the red lines shown in Fig. 6(b). The SNR of this ultrafast pulse is only 85.91%. This is because under these conditions, the gain length of main current peak is ~1.01 m, while the gain length of side current peak is ~1.25 m. After passing through the 17 m undulator, the radiation power of the main peak will be approximately ~26 times higher than that of the side peak. However, this contrast is insufficient for generating isolated pulses, which negatively affects the time-resolved scientific experiments. Overall, the advantages of our proposed scheme are clear.

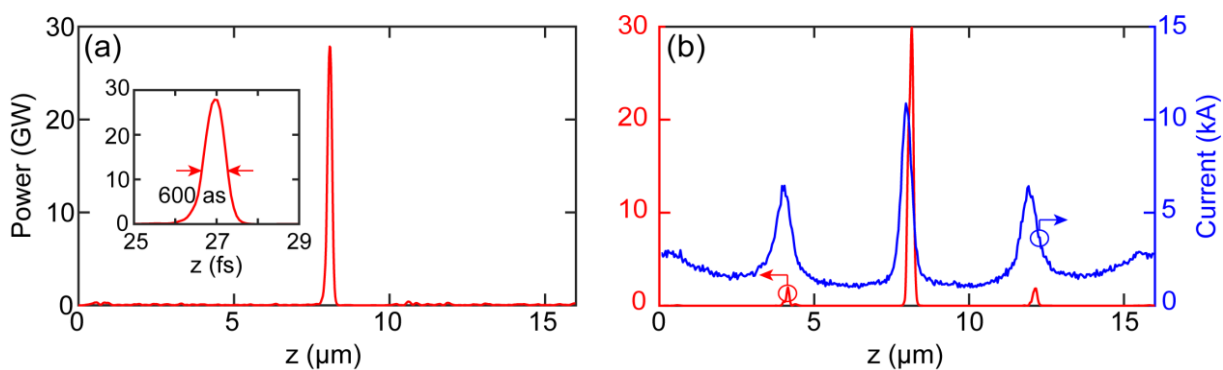

Fig. 6. The attosecond soft X-ray radiation at the output port of radiator for (a) our proposed scheme and (b) the conventional ESASE. In the right figure, the red curve represents the FEL pulse, and the blue curve represents the current distribution.

4. Conclusion

In conclusion, we proposed a novel scheme to produce high-intensity isolated attosecond FELs. Utilizing the sub-cycle MIR pulse generated through extreme soliton self-compression effect in an argon-filled HCF, one can introduce a quasi-isolated current peak into the electron beam, which in turn radiates an isolated ultrafast FEL pulse. In our simulation, the soft X-ray FEL pulses with duration of hundreds of attoseconds, peak power of tens of GW can be obtained at the output port of radiator. Most importantly, the generated isolated FEL pulse has an extremely high signal-to-noise ratio (about 96.4%). Such high-intensity ultrashort pulses can be used for numerous cutting-edge scientific applications, such as probing valence electron motion, photoemission delay, tunneling delay time and so on. In particular, by adopting the ultra-intense sub-cycle pulse synchronized with the attosecond FEL pulse as the pump laser, in pump-probe experiments, our proposed method will have broad application prospects.

The proposed method leverages advancements in traditional laser techniques and serves as an excellent research tool for attosecond X-ray spectroscopy. However, in practical experiments, potential issues such as laser jitter and slow drift may arise, causing fluctuations in attosecond FEL pulse width and intensity. Currently, the synchronization precision between the laser and electron beam can easily reach approximately 10 fs [53,54]. Since ESASE utilizes only a very short portion of the electron beam for lasing, and there is a uniform region of around 100 fs in the electron beam, the relative jitter and drift between the sub-cycle laser and the electron beam generally do not significantly affect the performance of the attosecond FEL. Nevertheless, feedback systems are required to control drift and ensure it remains within this range. It is also essential to carefully consider the effects of the environment and vibrations during experiments and to employ various feedback mechanisms to maintain the stability of the FEL output.

**Acknowledgements**

The authors thank Zhen Wang for helpful discussions about the calculation of longitudinal space charge field, and thank the project supported by Shanghai Municipal Science and Technology Major Project.

**Funding**

This work was supported by the National Natural Science Foundation of China (Nos. 12435011，11905275, 11775294, 12122514, and 62205353), the Youth Innovation Promotion Association CAS, the National Postdoctoral Program for Innovative Talents (No. BX2021328), the China Postdoctoral Science Foundation (No. 2021M703325) and the CAS Project for Young Scientists in Basic Research (YSBR-115).

[50] S. Reiche, *GENESIS 1.3: A Fully 3D Time-Dependent FEL Simulation Code*, Nucl Instrum Methods Phys Res A **429**, 243 (1999).

[51] Li S, Driver T, Rosenberger P, et al., *Attosecond coherent electron motion in Auger-Meitner decay*, Science **375**, 6578 (2022).

[52] Li S, Lu L, Bhattacharyya S, et al., *Attosecond-pump attosecond-probe x-ray spectroscopy of liquid water*, Science **383**, 6687 (2024).

[53] Zhao Z T, Wang D, Chen J H, et al. *First lasing of an echo-enabled harmonic generation free-electron laser*, Nature Photonics **6**, 360 (2012).

[54] Feng C, Liu T, Chen S, et al. *Coherent and ultrashort soft x-ray pulses from echo-enabled harmonic cascade free-electron lasers*, Optica **9**(7), 785-791 (2022).

# Supplemental Materials:

# Isolated attosecond free-electron laser based on a sub-cycle driver from hollow capillary fibers

Yaozong Xiao[1, 2, †], Tiandao Chen[1, 3, †], Bo Liu[2], Zhiyuan Huang[3, *], Meng Pang[3], Yuxin Leng[3], Chao Feng[2, *]

[1]Zhangjiang Laboratory, Shanghai 201210, China

[2]Shanghai Advanced Research Institute, Chinese Academy of Sciences, Shanghai 201210, China

[3]State Key Laboratory of High Field Laser Physics and CAS Center for Excellence in Ultra-intense Laser Science, Shanghai Institute of Optics and Fine Mechanics (SIOM), Chinese Academy of Sciences (CAS), Shanghai 201800, China

*Corresponding author. Email: fengc@sari.ac.cn (C. F), huangzhiyuan@siom.ac.cn (Z. H)

†These authors contribute equally to the work

The feature of the sub-cycle pulse generated from gas-filled hollow capillary fibers (HCFs) would affect significantly the final attosecond soft X-ray free-electron laser (FEL) radiation at the exit of the radiator. In this supplementary note, we present the simulation results of sub-cycle pulses generated under different conditions and the attosecond FEL radiation driven by them. In the simulations, the parameters of the FEL process are identical to those described in the main text, including the electron beam parameters, undulator parameters, and FEL radiation wavelength, etc. These results serve to illustrate typical dependency of the attosecond FEL performance and sub-cycle pulse property.

**I. Ultrafast FEL pulses generation driven by 800 nm sub-cycle laser**

Firstly, we studied the FEL pulse performance driven by the sub-cycle pulses with comparatively short carrier wavelengths. Figures S1 (a)-(c) show a ~ 1 fs sub-cycle pulse with a central wavelength of 800 nm. This sub-cycle pulse originates from a 30-cm long, 100-μm core HCF filled with negative pressure gradient (6 bar He to vacuum) pumped by 800 nm, 12 fs, 60 μJ few-cycle pulse. As shown in Figure S1 (d), such short-wavelength sub-cycle pulse gives rise to a quite narrow current spike. However, for the 1 nm soft X-ray FEL radiation, due to the slippage effect, excessively short current spikes can cause the radiation to rapidly slip out of the high-current region, affecting the generation of ultrafast pulses. As a result, the soft x-ray radiation at the exit of the radiator features multi-pulse SASE radiation rather than isolated attosecond pulse, see Figure S1 (e).

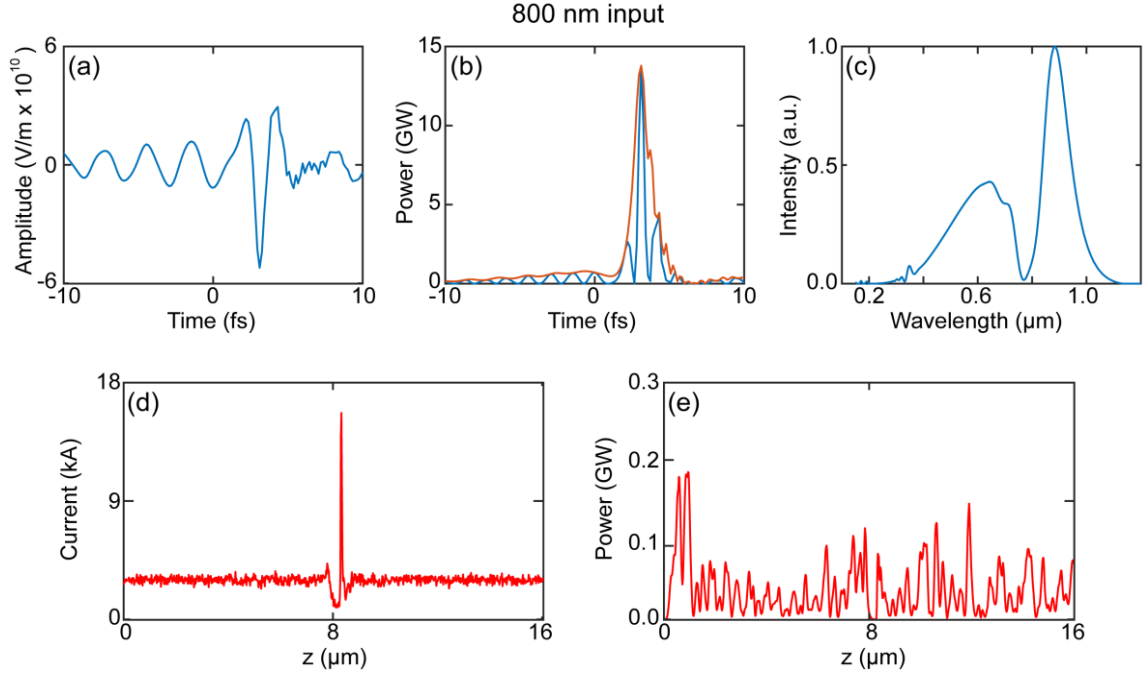

Figure S1. (a)-(c) Sub-cycle pulse generated through self-compression of a 12 fs, 800 nm, 60 μJ few-cycle pulse in a 30-cm-long, 100-μm-core HCF with decreasing pressure gradient. The simulation results are obtained using 2D model and the electric field in fundamental mode are presented. (a) The electric field amplitude of the sub-cycle pulses. (b) The pulse envelope (brown line) and electric field (blue line). (c) The corresponding pulse spectrum. (d) The current distribution before the radiator. (e) The FEL output pulse at the exit of the radiator.

**II. Ultrafast FEL pulses generation driven by 2000 nm sub-cycle laser**

Figures S2 (a)-(c) show a ~ 2 fs sub-cycle pulse with a central wavelength of 2000 nm. This sub-cycle pulse originates from a 30-cm long, 200-μm core HCF filled with negative pressure gradient (30 bar Ne to vacuum) pumped by 2000 nm, 20 fs, 85 μJ few-cycle pulse. In this case, the soft x-ray FEL radiation at the exit of the radiator is an attosecond pulse accompanied by some smaller sub-pulses. See Figure S2 (d) and (e) for the current distribution and FEL output. Relatively speaking, neither the signal-to-noise ratio (SNR) nor peak power of the FEL pulse is high enough.

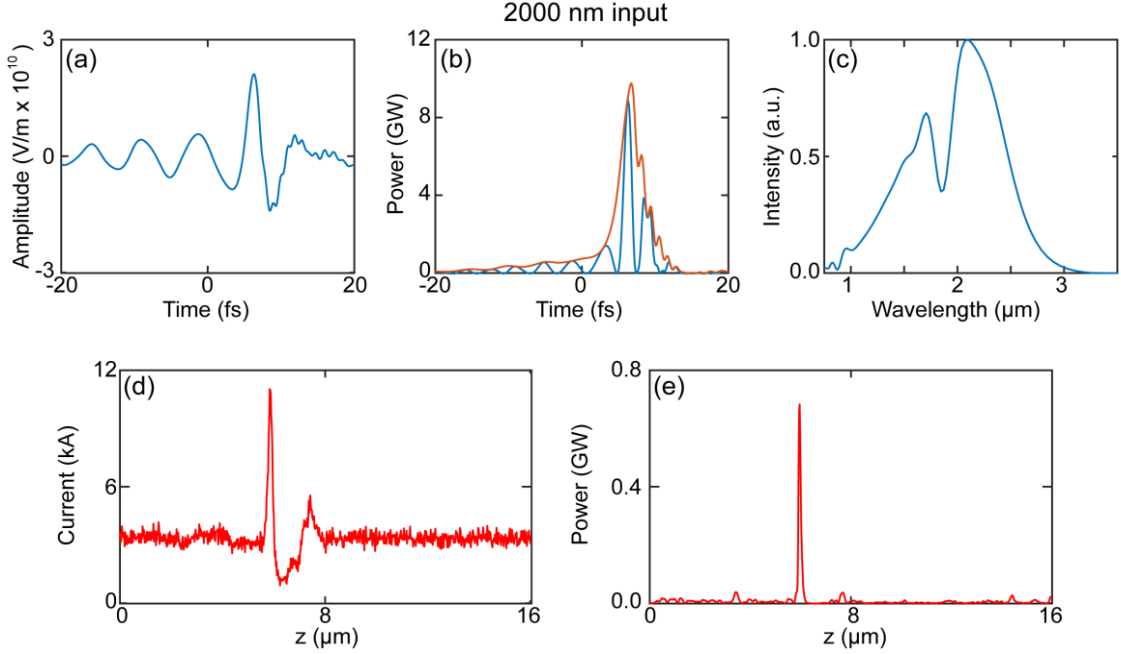

Figure S2. (a)-(c) Sub-cycle pulse generated through self-compression of a 20 fs, 2000 nm, 85 µJ few-cycle pulse in a 30-cm-long, 200-µm-core HCF with decreasing pressure gradient. The simulation results are obtained using 2D model and the electric field in fundamental mode are presented. (a) The electric field amplitude of the sub-cycle pulses. (b) The pulse envelope (brown line) and electric field (blue line). (c) The corresponding pulse spectrum. (d) The current distribution before the radiator. (e) The FEL output pulse at the exit of the radiator.

**III. Ultrafast FEL pulses generation driven by 4000 nm sub-cycle laser with side peaks**

The driving laser waveform inside the sub-cycle pulse profile will affect the attosecond soft X-ray FEL pulse. As an example, we present simulation results of two attosecond FEL pulses driven by two different sub-cycle pulses in Figures S4 and S5. The two driving pulses are mirror images of each other. Figures S3 (a)-(c) show a ~ 6 fs sub-cycle pulse with a central wavelength of 4000 nm, this sub-cycle pulse originates from a 30-cm long, 300-µm core HCF filled with negative pressure gradient (30 bar Ar to vacuum) pumped by 4000 nm, 40 fs, 55 µJ few-cycle pulse. As a comparison, Figures S4 (a)-(c) show the reverse version of the sub-cycle pulse in Figure S3. In Figure S3(d), we observe an additional current spike in the electron beam, which leads to an ultrafast FEL pulse with a clear side-peak pulse as shown in Figure S3(e). Meanwhile, in the Figure S4 case, the FEL pulse does not have any side-peak pulse, as shown in Figure S4(e).

The above phenomena are related to the magnetic compression process of electron beam in chicane (see Figure 2 in the main text). During passing through chicane, electrons with higher energy travel faster due to a shorter path than electrons with lower energy, enabling the magnetic compression process. Different electric field waveforms will introduce different energy chirps into the electron beam. Because of this, the two driving lasers, which are mirror images of each other, introduce completely different current modulation in the electron beam in Figure S3(d) and S4(d). For our proposed scheme, the presence of a relatively large side peak in the head of the sub-cycle laser as shown in Figure S4(b) is acceptable.

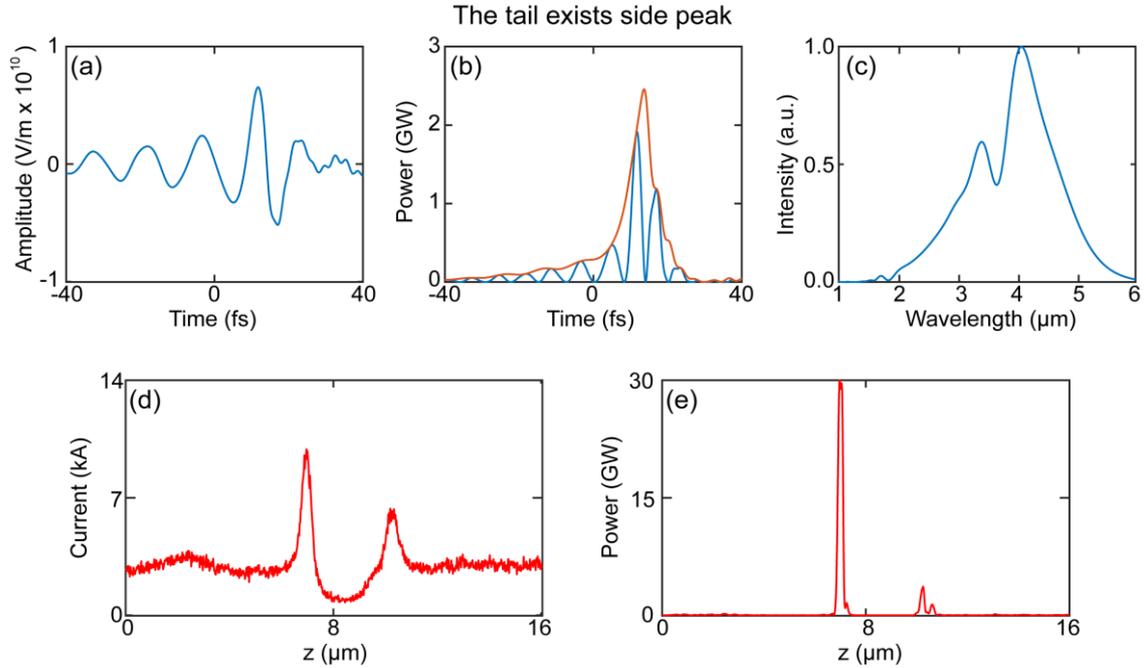

Figure S3. (a)-(c) Sub-cycle pulse generated through self-compression of a 40 fs, 4000 nm, 55 µJ few-cycle pulse in a 30-cm-long, 300-µm-core HCF with decreasing pressure gradient. The simulation results are obtained using 2D model and the electric field in fundamental mode are presented. (a) The electric field amplitude of the sub-cycle pulses. (b) The pulse envelope (brown line) and electric field (blue line). (c) The corresponding pulse spectrum. (d) The current distribution before the radiator. (e) The FEL output pulse at the exit of the radiator.

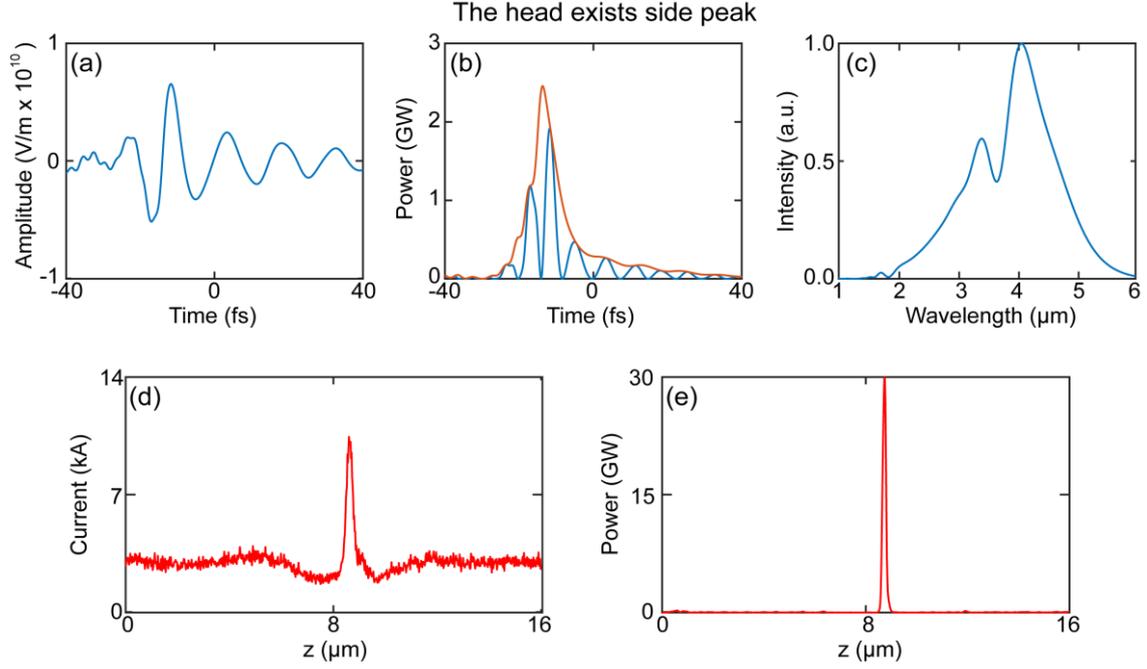

Figure S4. (a)-(c) Sub-cycle pulse generated through self-compression of a 40 fs, 4000 nm, 55 µJ few-cycle pulse in a 30-cm-long, 300-µm-core HCF with decreasing pressure gradient. The simulation results are obtained using 2D model and the electric field in fundamental mode are presented. (a) The electric field amplitude of the sub-cycle pulses. (b) The pulse envelope (brown line) and electric field (blue line). (c) The corresponding pulse spectrum. (d) The current distribution before the radiator. (e) The FEL output pulse at the exit of the radiator.

**IV. Ultrafast FEL pulses generation driven by 4000 nm sub-cycle laser with rapid oscillation at the trailing edge**

If the maximum self-compression point precedes the HCF exit, the phenomenon of phase-matched dispersive wave emission can be observed in subsequent propagation. This leads to fast varying modulation on the trailing edge of the output sub-cycle pulse. Surprisingly, the attosecond FEL radiation driven by such sub-cycle pulse still features high contrast. Figures S5 (a)-(c) show a ~ 3 fs sub-cycle pulse with a central wavelength of 4000 nm, this sub-cycle pulse originates from a 30-cm long, 300-µm core HCF filled with negative pressure gradient (113 bar He to vacuum) pumped by 4000 nm, 40 fs, 470 µJ few-cycle pulse. In Figures S6 (a) and (b), we observe the fast oscillation structure of the sub-cycle pulse, this structure corresponds to the spectral spike at short wavelength shown in Figure S5 (c). The current of the electron beam modulated by sub-cycle pulse have similar structure, however, we do not recognize this oscillation structure in the profile of the attosecond soft X-ray FEL radiation. See Figure S5 (d) and (e).

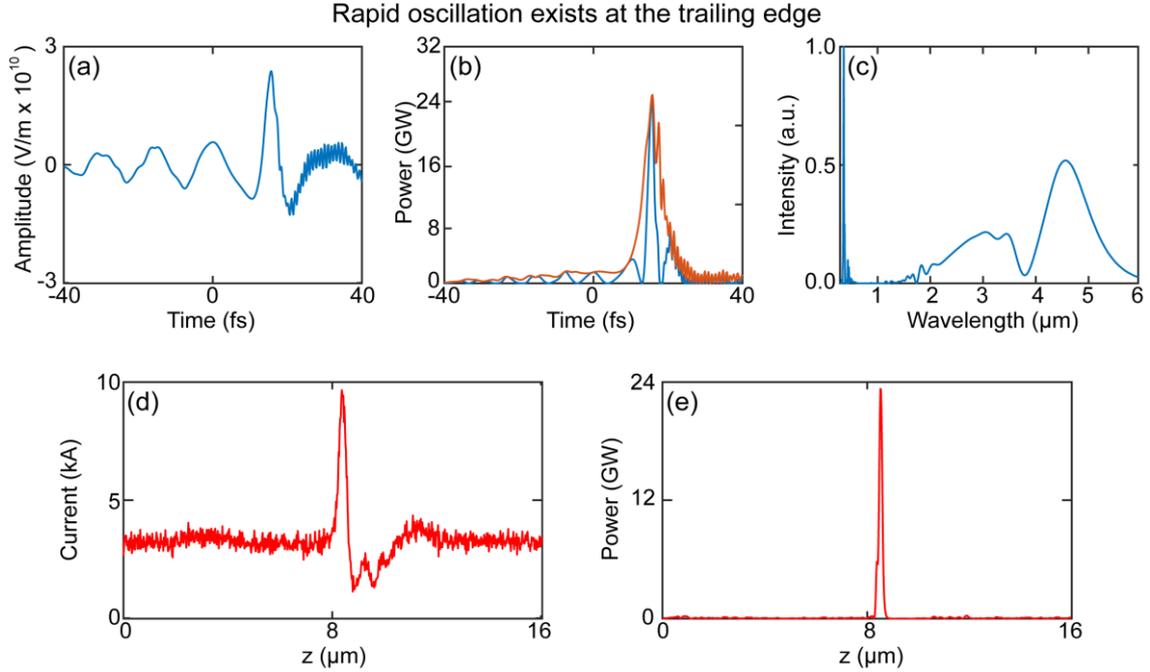

Figure S5. (a)-(c) Sub-cycle pulse generated through self-compression of a 40 fs, 4000 nm, 470 µJ few-cycle pulse in a 30-cm-long, 300-µm-core HCF with decreasing pressure gradient. The simulation results are obtained using 2D model and the electric field in fundamental mode are presented. (a) The electric field amplitude of the sub-cycle pulses. (b) The pulse envelope (brown line) and electric field (blue line). (c) The corresponding pulse spectrum. (d) The current distribution before the radiator. (e) The FEL output pulse at the exit of the radiator.

**V. Ultrafast FEL pulses generation driven by 4000 nm sub-cycle laser generated in Ar-filled HCF**

If the peak intensity of the pulse can exceed the ionization potential of the gas, the significant mode coupling can occur, and the phenomenon of blue shift soliton can be observed in propagation. This leads to a spectral blue-shift of the output sub-cycle pulse. Consequently, the attosecond FEL radiation driven by such sub-cycle pulse features comparatively low contrast. Figures S6 (a)-(c) show a ~ 2 fs sub-cycle pulse with a central wavelength of ~2000 nm, this sub-cycle pulse originates from a 30-cm long, 300-μm core HCF filled with negative pressure gradient (5 bar Ar to vacuum) pumped by 4000 nm, 40 fs, 420 μJ few-cycle pulse. In Figures S6 (a) and (b), we observe that the optical period of the sub-cycle pulse is shorter than that in Figure S5 case, corresponds to the spectral blue-shift shown in Figure S6 (c). Therefore, similar to the case of Figure S2, the attosecond FEL pulse from the radiator is accompanied by sub-pulses, see Figures S6 (d) and (e).

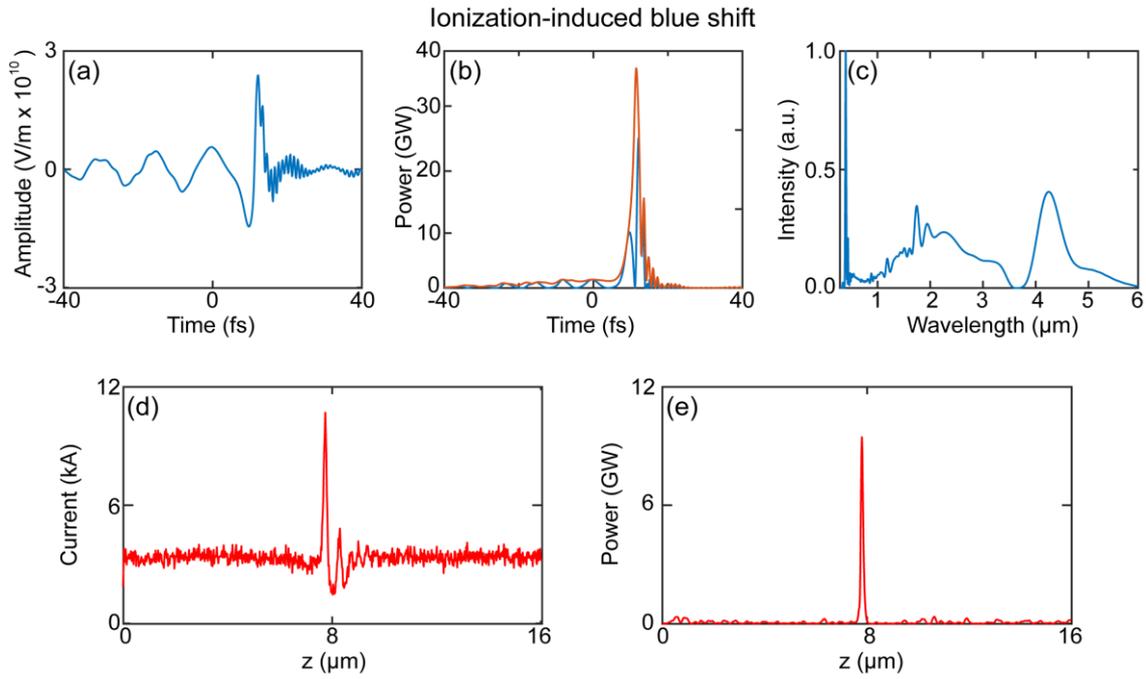

Figure S6. (a)-(c) Sub-cycle pulse generated through self-compression of a 40 fs, 4000 nm, 420 μJ few-cycle pulse in a 30-cm-long, 300-μm-core Ar-filled HCF with decreasing pressure gradient. The simulation results are obtained using 2D model and the electric field in fundamental mode are presented. (a) The electric field amplitude of the sub-cycle pulses. (b) The pulse envelope (brown line) and electric field (blue line). (c) The corresponding pulse spectrum. (d) The current distribution before the radiator. (e) The FEL output pulse at the exit of the radiator.

**VI. Influence of pump pulse energy fluctuation on ultrafast FEL pulses generation**

We found that both the sub-cycle pulses from HCF and the FEL pulses from radiator are insensitive to the small energy fluctuations of pump pulse, showing full robustness. As illustrate in Figures S7(a) and S7(b), when the energy of pump pulse increases or decreases by ~1%, the generated sub-cycle pulse is nearly unchanged. Accordingly, the FEL pulse driven by the sub-cycle pulse barely changed, only the peak power has fluctuated by ~3%, see Figure S7(d).

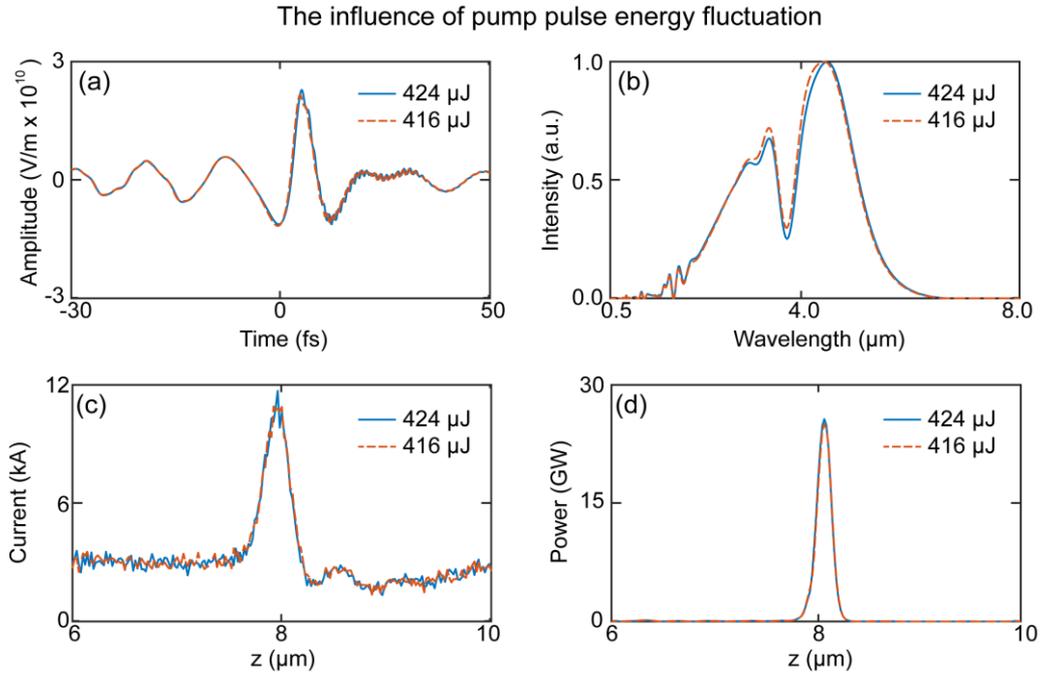

Figure S7. (a-b) Sub-cycle pulses generated through self-compression of 40 fs, 4000 nm, 416 µJ or 424 µJ few-cycle pulses in a 30-cm-long, 300-µm-core HCF with decreasing pressure gradient. The simulation results are obtained using 2D model and the electric field in fundamental mode are presented. (a) The electric field amplitude of the sub-cycle pulses. (b) The corresponding pulse spectra. (d) The current distribution before the radiator. (e) The FEL output pulse at the exit of the radiator.